\documentclass[showpacs, preprintnumbers,
nofootinbib, aps, prd, superscriptaddress,
10pt, showkeys, notitlepage,
twocolumn]{revtex4-1}
\usepackage{graphicx,amssymb,amsmath,amsthm,amsfonts,epsfig,natbib}
\usepackage[utf8]{inputenc}
\usepackage[linktocpage,breaklinks]{hyperref}
\usepackage[usenames,dvipsnames]{color}
\usepackage{tikzsymbols}
\usepackage{epstopdf}
\usepackage{aas_macros}
\usepackage{tensor}
\usepackage{mathtools}
\usepackage{amsbsy}
\usepackage{bm}

\definecolor{darkred}{rgb}{0.5,0,0}
\definecolor{darkgreen}{rgb}{0,0.5,0}
\definecolor{darkblue}{rgb}{0,0,0.5}
\definecolor{prussian}{rgb}{0.0, 0.19, 0.33}
\definecolor{richelectricblue}{rgb}{0.03, 0.57, 0.82}
\definecolor{teal}{rgb}{0.0, 0.5, 0.5}
\definecolor{mediumseagreen}{rgb}{0.24, 0.7, 0.44}
\definecolor{lust}{rgb}{0.9, 0.13, 0.13}
\definecolor{ballblue}{rgb}{0.13, 0.67, 0.8}
\definecolor{darkcyan}{rgb}{0.0, 0.55, 0.55}
\definecolor{mountainmeadow}{rgb}{0.19, 0.73, 0.56}
\definecolor{palecarmine}{rgb}{0.69, 0.25, 0.21}
\definecolor{richcarmine}{rgb}{0.84, 0.0, 0.25}
\definecolor{tangelo}{rgb}{0.98, 0.3, 0.0}
\definecolor{venetian}{rgb}{0.784,0.031,0.082}
\definecolor{bdfrance}{rgb}{0.192,0.549,0.906}

\hypersetup{colorlinks=true, citecolor=darkblue, linkcolor=darkblue,
urlcolor = darkblue}

\newcommand{\be}{\begin{equation}}
\newcommand{\ee}{\end{equation}}
\newcommand{\bear}{\begin{eqnarray}}
\newcommand{\eear}{\end{eqnarray}}
\newcommand{\ba}{\[\begin{aligned}}
\newcommand{\ea}{\end{aligned}\]}
\newcommand{\nn}{\nonumber}

\newcommand{\p}{\prime}
\newcommand{\pp}{\prime\prime}

\newcommand{\Om}{{\Omega}}

\newcommand{\rK}{{\rm K}}
\newcommand{\rR}{{\rm R}}

\newcommand{\rph}{{\rm ph}}

\begin{document}
\title{How well can ultracompact bodies imitate black hole ringdowns?}

\begin{abstract}
The ongoing observations of merging black holes by the instruments of the fledging gravitational wave astronomy 
has opened the way for testing the general relativistic Kerr black hole metric and, at the same time, for probing the existence 
of more speculative horizonless ultracompact objects. In this paper we quantify the difference that these two classes of objects 
may exhibit in the post-merger ringdown signal. By considering rotating systems in general relativity and assuming an eikonal
limit and a third-order Hartle-Thorne slow rotation approximation, we provide the first calculation of the early ringdown 
frequency and damping time as a function of the body's multipolar structure. Using the example 
of a gravastar, we show that the main ringdown signal may differ as much as a few percent with respect to that of a
Kerr black hole, a deviation that could be probed by near future Advanced LIGO/Virgo searches. 

\end{abstract}

\author{Kostas Glampedakis}
\email{kostas@um.es}
\affiliation{Departamento de F\'isica, Universidad de Murcia, Murcia, E-30100, Spain}
\affiliation{Theoretical Astrophysics, University of T\"ubingen, Auf der Morgenstelle 10, T\"ubingen, D-72076, Germany}

\author{George Pappas}
\email{georgiospappasgr@gmail.com}
\affiliation{Department of Physics and Astronomy,
The University of Mississippi, University, MS 38677, USA}
\affiliation{School of Mathematical Sciences, University of Nottingham, University Park, Nottingham NG7 2RD, UK}
\affiliation{Departamento de F\'isica, CENTRA, Instituto Superior T\'ecnico,
Universidade de Lisboa, Avenida Rovisco Pais 1, 1049 Lisboa, Portugal}


\maketitle


\section{Introduction}
With the direct observation of merging black hole binaries by the LIGO detectors
~\cite{Abbott:2016blz, Abbott:2016nmj, Abbott:2017vtc,GW170814arxiv}, the last two years saw the metamorphosis  of 
gravitational waves (GWs) from a mostly theoretical concept to a tangible astrophysical tool. With many more 
detections to come in the near future, unprecedented precision tests of general relativity (GR) are 
already within reach~\citep{Yunes:2013dva,Berti:2015itd,TheLIGOScientific:2016src}.  

One of the most exciting prospects in this context is to probe the Kerr spacetime of rotating black 
holes as a unique prediction of GR.  This signature property of the theory can be tested
by observations of the final ringdown signal of the merger remnant and the extraction of its quasi-normal mode
(QNM) frequencies and damping times (see~\cite{Kokkotas:1999bd,Berti:2009kk} for a review), a ``black hole spectroscopy" 
technique \cite{Detweiler:1980gk,Dreyer:2003bv,Berti:2005ys} akin to atomic line spectra. 

Over the years, however, theorists have proposed a number of alternative GR models of ultracompact objects (UCOs) 
that could be easily mistaken (by ordinary photon astronomy) for black holes while in fact they
are horizonless and made of some kind of exotic matter (such as scalar fields or anisotropic/negative pressure fluid).
Loosely speaking, these objects are defined as being compact enough as to allow the Schwarzschild photon ring radius 
($r_{\rm ph} =3M$ in geometric units $c=G=1$, where $M$ is the mass) to lie outside the body's surface.  
Ordinary neutron stars clearly fail to meet this requirement (nor are massive enough for that matter) but exotic objects 
such as gravastars, bosons stars, etcetera  fit the bill more than enough (see~\cite{Cardoso_Pani:2017} for a recent review). 

There is an intimate connection between the photon ring (which is roughly located at  the maximum of the
exterior spacetime wave potential) and the fundamental QNM. Indeed, a black hole ringdown (as the one 
observed in the first detection GW150914~\cite{Abbott:2016blz}) is dominated by the fundamental mode and physically 
represents radiation backscattered at the photon ring/potential peak.  
Recent work~\cite{Cardoso:2016rao, Cardoso:2016oxy, Mark:2017zmc} uses the photon ring/QNM link to claim 
indistinguishability between \emph{non-rotating} black holes and UCOs.  Although these two classes of systems are known 
to support markedly different QNM spectra, they nevertheless display similar ringdown waveforms -- this is a consequence of 
these objects sharing the same Schwarzschild wave potential/photon ring in the exterior spacetime as enforced by Birkhoff's 
theorem (see \cite{Nollert:1996rf} for an early discussion on the subject). 

The similarity in the black hole/UCO ringdown could be a potentially serious problem for GW-based tests 
of GR, but there are at least two ways to lift the degeneracy.
First, UCOs are known to support  slowly damped $w$-modes in the ``cavity" formed 
between the wave potential's peak and the body's center/surface~\cite{1991RSPSA.434..449C, Kokkotas:1994an}. 
These modes show up at a later part of the signal in the form of ``echoes"~\cite{Mark:2017zmc,Maselli_etal2017}, 
a feature that is absent in the ringdown of garden-variety black holes.
In fact, the $w$-modes may even modify the ringdown from the very beginning, as is the case for 
compact gravastars~\cite{Chirenti_Rezzolla:2007,Chirenti_Rezzolla:2016}.
Second, the astrophysical existence of UCOs continues to be questioned and to attract strong criticism 
due to the lack of realistic formation mechanisms and the intrinsic instabilities that these objects may 
suffer~\cite{1978MNRAS.182...69S, Kokkotas:2002sf,Cardoso:2007az,Keir:2014oka, Cunha_etal2017}.

In this paper we revisit the topic of black hole/UCO ringdown indistinguishability by considering \emph{rotating}
systems which are not constrained by Birkhoff's  theorem to share the same exterior spacetime structure and
thus could provide another observational ``handle" for discerning the nature of ultracompact objects. 
Adopting a somewhat empirical attitude, we focus on the  early (and strongest) part of the ringdown signal 
associated with the photon ring while we remain agnostic about the presence of late-time echoes or the stability of the UCO itself.  
By means of a slow-rotation expansion, and working within the framework of GR, we obtain the exterior spacetime metric 
associated with an arbitrary stationary-axisymmetric UCO as a function of the mass, spin and higher multipole moments.
This metric is then used to calculate the photon ring properties (orbital frequency, Lyapunov decay rate). 
Knowledge of these parameters allows the calculation of the ringdown signal's frequency and damping rate
within the eikonal limit approximation, a well-established technique in the context of Kerr black holes
~\cite{Goebel:1972,Ferrari:1984zz,Kokkotas:1999bd,Berti:2009kk,Cardoso:2008bp,Dolan:2010wr,Yang:2012he,Yang:2012pj,Yang:2013uba}.
As an illustrative example, we consider a thin-shell gravastar and compare its approximate ringdown against that
of a Kerr black hole.


\section{Constructing the UCO spacetime}
A rotating UCO is assumed to be a stationary and axisymmetric body and therefore its spacetime structure 
can be described as (here we adopt spherical coordinates $\{r,\theta,\varphi\}$),
\begin{align}
ds^2 &=  -e^\nu \left [ 1 + 2 \epsilon^2 \left (  h_0  + h_2 P_2\right )  \right ] dt^2 
\nn \\
&+ \left (1 -\frac{2m}{r} \right )^{-1} \left [ 1 +  2 \epsilon^2 \frac{(\mu_0  + \mu_2  P_2 )}{r-2m} \right ]dr^2 
\nn \\
& + r^2 \left ( 1 + 2 \epsilon^2  k_2 P_2 \right ) \Bigg [ d\theta^2 + \sin^2\theta \left [   d\varphi  \right.
\nn \\
 & \left.  -   \epsilon \left \{ \Om -\omega_1 P_1 + \epsilon^2 \left  ( w_1 P_1^\p + w_3  P_3^\p \right ) \right \} dt  \right  ]^2 \Bigg ],
\label{metric}
\end{align}
where $P_\ell (\cos\theta)$ is the standard Legendre polynomial, $\{\nu,m,\mu_0,\mu_2,h_0,
h_2,k_2,\omega_1,w_1,w_3\}$ are radial functions and a prime represents a derivative with respect to 
the argument. With respect to the body's angular frequency $\Om$, (\ref{metric}) is the $\mathcal{O} (\Om^3)$ 
extension of the classic Hartle-Thorne metric \cite{Hartle1967,HT68}, with the bookeeping parameter $\epsilon$ indicating the 
$\Omega$-order of each term. The functions $\nu, m$ give rise to the `background'  spherically 
symmetric star with radius $R$ and mass $M \equiv m(r>R)$, while the rest of the functions are the induced rotational 
perturbations.

Once the above metric is inserted into the $\mathcal{O} (\Om^3)$ truncated GR field equations, the 
resulting expressions can be solved analytically in the body's vacuum exterior without having to appeal 
to the specific properties of the interior matter. For the purpose of the present work it is enough to show the 
solutions for $\omega_1,w_1, w_3$, the rest of the solutions can be found in the literature, see e.g. \cite{Hartle1967,Pani2015}. 
In terms of the dimensionless parameters $x = r/M$  and $\chi = S/M^2$, where $S\sim \mathcal{O} (\Om)$ is 
the angular momentum, we have $ \omega_1 =   \Omega -  2 \chi/ M x^3 $ and
\begin{align}
w_1 & = \frac{\chi^3}{10Mx^6} \left [\,  -8 \left  (1+\frac{3}{x} \right ) + 20 x^3\, \delta s  \right.
\nn \\
& \left. +  \frac{5}{4}\delta q x^2 \Bigg \{\,  3x (x-2)(5x^2 -2x-4) \log \left (1-\frac{2}{x} \right )  \right.
\nn \\
& \left. \quad   +\, 2 ( 15x^3 -21x^2 -16x +6)  \,\Bigg \}  \, \right ],
\label{w1_ext}
\end{align}
\begin{align}
& w_3  = - \frac{\chi^3}{15 M x^5} \Bigg [\,  2 \left (5+\frac{9}{x} \right ) \left ( 1 -\frac{2}{x} \right ) 
\nn \\
& - \frac{15}{16} x^2 \delta s_3 \Bigg \{ \frac{35}{2} x^3 (3x-4) (x-2)  \log \left (1-\frac{2}{x} \right )  
\nn \\
&  \quad +\,  \frac{7}{3} (45x^4 -105x^3 +30x^2 +10x +4 )  \Bigg \}   
\nn \\
&  + \frac{5}{8}x \delta q \Bigg \{  315x^5-735x^4+ 210x^3 + 34x^2  + 52x +36 
\nn \\
& + \frac{3}{2} x(x-2) (105x^4 -140x^3 -12x -4) \log \left (1-\frac{2}{x} \right )  \Bigg \} \, \Bigg  ].
\label{w3_ext}
\end{align}
The exterior solutions contain a number of unspecified integration constants that can be cast in a dimensionless form
with clear physical meaning: $\delta m$ (appearing in metric functions not shown here) and $\delta s$ represent the 
rotational corrections to the body's mass and spin,
\be
\delta M = \chi^2 M \delta m, \quad \delta S = \chi^3 M^2 \delta s,
\ee
while $\delta q, \delta s_3$ describe, respectively, the deviation of the UCO's  Geroch-Hansen  mass quadrupole $M_2$ and 
spin octupole $S_3$ moments (note that these are the moments typically used in formulating  ``hair theorems" for compact
objects, see~\cite{Pappas:2012,Yagi_etal:2014,DonevaPappas2017} and references therein) from the 
corresponding moments of a Kerr black hole,
\be
M_2  =  -\chi^2 M^3 \left ( 1 - \delta q \right ), ~  S_3 =   -\chi^3 M^4 \left (1 - \delta s_3 \right ).
\ee
Eventually, all these constants are specified once the body's interior metric is calculated 
and matched to the exterior via the appropriate surface junction conditions.


\section{UCO ringdown properties}
The rigorous calculation of  QNM spectra and ringdown signals of (at least $\mathcal{O} (\Omega^2))$ rotating 
UCOs is still beyond reach due to the non-separability of the perturbed GR equations
(in fact such calculation is still lacking even for ordinary rotating neutron stars). 

Focusing on the ringdown signal associated with the potential peak/photon ring, the next best option is to work in
the short wavelength eikonal limit~\cite{Goebel:1972,Ferrari:1984zz,Kokkotas:1999bd,Berti:2009kk,Dolan:2010wr,
Yang:2012he,Yang:2012pj,Yang:2013uba} where the ringdown's frequency and damping rate are determined
 by the geodesic photon ring's angular frequency and Lyapunov exponent, respectively. 
For Kerr black holes, where the main ringdown is identified with the fundamental QNM, this approximation 
is known to work extremely well with a typical accuracy of a few percent, see discussion below and Fig.~\ref{fig:gravastar_vs_Kerr}.  
We expect the same to be true for UCOs although in that case the photon ring 
is related with the ringdown but not with the QNMs themselves~\cite{Cardoso:2016rao,Price:2017cjr}.

By means of the analytically known exterior metric, we can produce slow-rotation expressions for the spacetime's (prograde) 
photon ring radius $r_{\rm ph}$, angular frequency $\Omega_{\rm ph}$ and Lyapunov exponent $\gamma_{\rm ph}$. 
To that end, we first consider the  $r_\rph, \Omega_\rph, \gamma_{\rm ph}$ equations for a  general axisymmetric-stationary 
metric $g_{\mu\nu} (r,\theta)$ \cite{Glampedakis:2017}:
\begin{align}
& g_{\varphi\varphi}  (g^\prime_{tt} )^2  + 2 g_{tt} (g^\prime_{t\varphi} )^2 -g^\prime_{tt} \left (\, g_{tt} g_{\varphi\varphi}^\prime 
+ 2 g_{t\varphi} g_{t\varphi}^\prime   \,\right )
\nn \\
& - 2 \sqrt{W} \left  (\, g_{t\varphi} g^\prime_{tt} - g_{tt} g^\prime_{t\varphi} \, \right )    = 0,
\label{phring}
\\
&\Om_{\rm ph}=   - \frac{ g_{tt}^\prime}{ \sqrt{W} + g_{t\varphi}^\prime }, \quad 
W =  (g^\prime_{t\varphi})^2 - g_{tt}^\prime g_{\varphi\varphi}^\prime,
\label{Omph}
\\
&\gamma^2_{\rm ph}  = \frac{ \Om_{\rm ph}^2  \left ( g_{t\varphi}^2 - g_{tt} g_{\varphi\varphi} \right ) }
{2g_{rr}( g_{tt}  + \Om_{\rm ph} g_{t\varphi} )^2} 
\left ( g_{tt}^{\pp}  + 2 g_{t\varphi}^{\pp} \Om_{\rm ph}+ g_{\varphi\varphi}^{\pp} \Om_{\rm ph}^2 \right ),
\label{gamma_ph}
\end{align}
where 
$g_{\mu\nu} (r,\pi/2)$ is used for the equatorial photon ring. The eikonal  frequency  $\omega_{\rm R}$ and 
damping rate $\omega_{\rm I}$ for any ``equatorial" $\ell=m$ mode are given by 
$\omega_\rR = \ell\, \Omega_{\rm ph},~\omega_{\rm I} = |\gamma_{\rm ph}|/2$, 
with $\ell=2$ being the dominant quadrupole mode. 

Inserting the exterior metric in Eqs. (\ref{phring}), (\ref{Omph}) and keeping 
terms up to ${\cal O}(\Om^3)$ we obtain the photon ring radius,
\begin{align}
&x_{\rm ph} = 3 -\frac{2\chi}{\sqrt{3}}  -  \chi^2\left [\, \frac{1}{27} - 3 \delta m  + \frac{5}{4} \left ( 13 - \frac{45}{4} \log 3  \right ) \delta q  \, \right ]
\nn \\ 
& -  \frac{\sqrt{3}}{2}  \chi^3  \left [\,  \frac{4}{81} \left ( 1 + 27 \delta s - 27 \delta m  \right ) -  \left ( 553 -505 \log 3  \right )  \delta q   \right.
\nn \\
& \left.  + \, \frac{7}{4} \left (   148 - 135 \log 3  \right ) \delta s_3    \, \right ], 
\label{xLR}
\end{align}
the $\ell=2$  ringdown frequency,
\begin{align}
& M \omega_\rR = \frac{2}{3\sqrt{3}} + \frac{4\chi}{27} + \frac{2\chi^2}{3\sqrt{3}}  
\Bigg [ \frac{11}{54}  - \delta m  +  5 \left (  \frac{15}{16} \log3 -  1\right ) \delta q  \Bigg   ]
\nn \\
& + \frac{4 \chi^3}{81}  \Bigg [\,  1 -9\delta m + 3 \delta s  + \frac{9}{64} \left (\, 5652 -5125\log3 \, \right ) \delta q  
\nn \\
& -\frac{21}{128} \left (\, 2228 -2025\log3\,\right) \delta s_3  \, \Bigg ],
\label{omR}
\end{align}
and the damping rate,
\begin{align} 
& M \omega_{\rm I} =\frac{1}{6 \sqrt{3}} - \frac{\chi ^2}{6\sqrt{3}} \left [ \frac{4}{54} + \delta m 
+ \frac{5}{8}  \left(15 \log 3 -16\right) \delta q \right ]
\nn
\\
& - \frac{\chi ^3}{93312} \Bigg [\,  640 -  270 \left (14595 \log 3-16076 \right) \delta q 
\nn \\
& + 945  \left(2025 \log 3 -2228\right ) \delta s_3 \, \Bigg ],
\label{omI}
\end{align}   
as functions of the UCO's multipolar structure.

Eqs.\,(\ref{omR})-(\ref{omI}) represent the central result of this paper.
As far as we are aware these are the most general expressions in the literature
describing the ringdown associated with the photon ring of an arbitrary
axisymmetric and stationary UCO in GR.
They can be compared against 
the eikonal-limit Kerr QNM/ringdown frequency and damping rate, expanded to third order with respect to the 
dimensionless Kerr parameter $\bar{a} = S_{\rm BH}/M^2_{\rm BH}$, with $S_{\rm BH}$ and 
$M_{\rm BH}$ denoting, respectively, the black hole's total spin and gravitational mass:
\begin{align}
M_{\rm BH} \omega_\rR^\rK &= \frac{2}{3\sqrt{3}} + \frac{4}{27} \bar{a} + \frac{11}{81\sqrt{3}} \bar{a}^2 + \frac{4}{81} \bar{a}^3,
\label{omR_a3}
\\
M_{\rm BH} \omega_{\rm I}^\rK &=  \frac{1}{6\sqrt{3}} - \frac{1}{81\sqrt{3}} \bar{a}^2 - \frac{5}{729} \bar{a}^3.
\label{omI_a3}
\end{align}
In order to make a meaningful comparison with the UCO expressions they too ought to be written
in terms of the \emph{total} mass $M_{\rm tot} = M (1+ \chi^2 \delta m) $ and spin $ S_{\rm tot}= \chi M^2 ( 1+ \chi^2 \delta s )$. 
Crucially, this change is imposed by the identification of $M_{\rm tot}, S_{\rm tot}$ as the body's observable
mass and spin. Inspection of (\ref{xLR})-(\ref{omI}) reveals that changing 
\be
M \to M_{\rm tot}, ~ \chi \to \chi_{\rm tot} = \frac{S_{\rm tot}}{M^2_{\rm tot}} = \chi + \chi^3 ( \delta s -2 \delta m), 
\label{newparams}
\ee
amounts to setting $\delta m = \delta s =0$ in these equations since the two parameters 
are completely ``absorbed" in the total mass and spin. Once this is done, the modified expansions 
(\ref{omR}) and (\ref{omI}) allow a key conclusion to be drawn: the photon ring-related ringdown 
of a rotating UCO does \emph{not} match that of a Kerr black hole, the difference originating
from the multipolar deviations  $\{\delta q,\delta s_3\}$. To make this statement quantitative
we will have to consider a particular example of UCO.


\section{Application: thin-shell gravastar}
As an illustrative application of the UCO formalism we consider the so-called 
thin-shell gravastar model \cite{Visser:2004}. This is a `star'  consisting of a de Sitter core 
and a fluid surface layer. The gravastar's key property lies in its ability to have a compactness 
$C = M/R$ arbitrarily close to that of a Schwarzschild black hole, $C_{\rm BH} = 1/2$. 
Although a gravastar's ringdown might be contaminated early on by 
its $w$-modes~\cite{Chirenti_Rezzolla:2016}, it is an ideal benchmark for the kind
of physics we are interested in here.

A $\mathcal{O} (\Omega^2)$ rotating  gravastar with a thin shell of vanishing 
density and finite surface tension has been the subject of recent work~\cite{Pani2015}. 
This model allows an analytic calculation of the interior metric  that can be smoothly joined across 
the gravastar surface to the exterior one. 
Following Ref.~\cite{Pani2015} we obtain the quadrupole term $\delta q$ as a 
function of  $C$,
\begin{align}
\delta q  &=  \frac{5}{4} \frac{C^5}{ \Delta_0} \Bigg [\,  2  \sqrt{2 C} ( 16C^2 -6C -9)  
\nn \\
&  +  9 (1- 4C^2) \log \left ( \frac{1+\sqrt{2C}}{1-\sqrt{2C}} \right ) \, \Bigg ],
\label{dqG}
\end{align}
where
\begin{align}
\Delta_0  &= \sqrt{2 C} \left [\, 2C (8C^3 + 9C^2 -12C +9)  \right.
\nn\\
& \left. + 3 (6C^2 -7C +3) \log (1-2C) \, \right ] 
\nn \\
& - \frac{3}{2}  \log \left ( \frac{1+\sqrt{2C}}{1-\sqrt{2C}} \right )  \Bigg [\,  2C (6C^3 -5C^2 -6C +3) 
\nn \\
&  +  3(4C^3 -3C +1) \log(1-2C) \, \Bigg ].
\end{align}
One also finds $\omega_1 (r<R) =0$ which leads to $\Omega = 2S/R^3$.

In order to derive similar expressions for the remaining constants $\delta s, \delta s_3$ we need to 
add $\mathcal{O} (\Omega^3)$ physics to the above gravastar model. The interior equations for 
$w_1, w_3$ can be found using the general formalism of Ref.~\cite{Benhar_etal2005}:
\begin{align}
&w_1^{\pp} + \frac{4}{r} w_1^\p = 0,  
\\
&w_3^{\pp} +  \frac{4}{r}  w_3^\p  - \frac{10}{r^2} \left (1-\frac{2M r^2}{R^3} \right )^{-1} w_3=  0.
\end{align}
A particular characteristic of the gravastar's structure is that these equations take a homogeneous form, 
completely decoupling themselves from the rest of the metric functions (unlike what would be the case
for, say, a normal rotating neutron star). This implies that we need to set $w_1 = w_3 =0$ in the gravastar 
interior in order to be able to determine the exterior constants $\delta s, \delta s_3$. This requires
the exterior solutions (\ref{w1_ext}), (\ref{w3_ext}) to vanish at the surface. We then find:
\begin{align}
\delta s    & =  -\frac{\delta q}{20 C^3} \left [  \, 2C ( 6C^3 -16C^2 -21C +15)  \right.
\nn \\
& \left. +  3 (  8C^3 -12C + 5 ) \log (1-2C) \, \right ]  + \frac{2}{5}   C^3 (1+3C),
\label{dsG}
\\
\delta s_3  &=  \frac{2}{35} \left [\,  (1-2C) \frac{\Delta_1}{\Delta_2}  +  \frac{115}{4} \delta q \, \right ],
\label{ds3G}
\end{align}
where
\begin{align}
&\Delta_1  =  32 C^7 (5+9C)   -\delta q  \left [ \, 2C(72C^4 +94 C^3 + 75 C) \right.
\nn \\
& \left.   +\, 3\log(1-2C) (16C^4 + 48C^3 +100C -75)\, \right ],
\\
&\Delta_2  =  2C( 4C^4 +10C^3 +30C^2 -105C +45) 
\nn \\
& \qquad   + 15 (1-2C)(3-4C) \log(1-2C).
\end{align}
Eqs.~(\ref{dqG}), (\ref{ds3G}) predict a monotonic decay  $\{\delta q, \delta s_3 \} \to 0 $ in the 
`BH limit' $C \to C_{\rm BH}$ and therefore the exterior spacetime of  the $\mathcal{O}(\Om^3)$ gravastar 
will increasingly look like a Kerr hole with increasing $C$, in agreement with the analysis of~\cite{YY2015} for 
anisotropic fluid stars.   

\begin{figure}[b]
\includegraphics[width=0.5\textwidth]{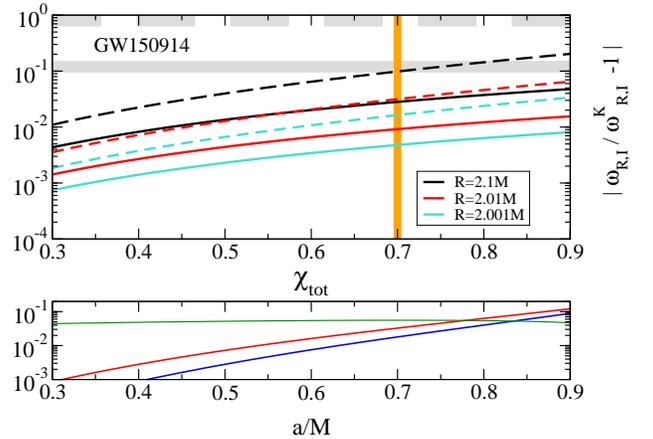}
\caption{\emph{Gravastar vs Kerr black hole ringdown.}
Top: the fractional difference in the ringdown frequency $\omega_\rR$ (solid curves) and
damping rate $\omega_{\rm I}$ (dashed curves) as a function of the total spin
$\chi_{\rm tot}$, Eq.~(\ref{newparams}), for a thin shell gravastar of radius
$ R/M= 2.1, 2.01, 2.001$. The vertical line marks the typical spin of the observed 
post-merger remnants while the thick horizontal bands represent the accuracy of the measured 
ringdown frequency (solid) and damping time (dashed) of  GW150914.
Bottom: the precision of the slow-rotation approximation is shown 
as  the fractional difference in the Kerr $\omega_\rR$ between each of the 
$\mathcal{O}(a^3) $ and $\mathcal{O}(a^4)$  expansions and the full eikonal result  
(middle and bottom curve, respectively). The top curve represents the fractional precision of the 
eikonal approximation with respect to the numerical Kerr $\omega_\rR$.}
\label{fig:gravastar_vs_Kerr}
\end{figure}

By combining the multipole moment results (\ref{dqG}), (\ref{ds3G}) with the general formulae 
(\ref{omR}), (\ref{omI}) we can quantify the difference in the ringdown between a thin gravastar 
and a Kerr black hole. The results are displayed in Fig.~\ref{fig:gravastar_vs_Kerr} in the form of 
fractional differences $ | \omega_{\rm R,I}/\omega^\rK_{\rm R,I}-1|$ as functions of $\chi_{\rm tot} = \bar{a}$ and for
different choices of $C$. We can see that for a wide range of parameters,
$ 2.001 \lesssim R/M \lesssim 2.1 $, and including the most relevant spin for merging binaries, 
$\chi_{\rm tot} \approx 0.7$, the differences in the frequency and damping are $\sim 0.5-2\%$ and 
$ \sim 2-10\%$ respectively. 
These numbers lie about an order of magnitude below the ringdown accuracy achieved in 
GW150914~\cite{TheLIGOScientific:2016src}. In terms of the deviation from the Kerr multipole moments, 
the previous numbers translate to $0.05 \lesssim \{ \delta q, \delta s_3\} \lesssim 0.3$. 

Fig.~\ref{fig:gravastar_vs_Kerr} also displays the precision associated with 
the two approximations underpinning this work, the slow-rotation expansion and the 
eikonal limit. Given the similarity between the UCO and Kerr  $\omega_\rR$
the latter can serve as a precision template (the error in $\omega_{\rm I}$ is similar and therefore not shown). 
The error associated with the eikonal limit is  $5-6\%$ and has been estimated by 
comparing the full eikonal expression (see \cite{Ferrari:1984zz} for details) against the exact numerical value of
the fundamental Kerr QNM (computed with the help of the fit formula of~\cite{Glampedakis:2017}). 
The fractional difference of the $\mathcal{O}(a^3)$ expansion (\ref{omR_a3}) with 
respect to the full eikonal result is  $\lesssim 6\%$. As this lies below the eikonal 
limit error for any spin $ a \lesssim 0.8M$, the use of slow-rotation is perfectly 
justified in the same spin range. These numbers also imply that the fractional difference between 
the $\mathcal{O}(a^3)$ eikonal expansion and the full Kerr QNM is essentially the same as
the precision of the full eikonal result. Moreover, Fig.~\ref{fig:gravastar_vs_Kerr} suggests that 
a more accurate $\mathcal{O}(a^4)$  eikonal expansion leads to a marginal improvement for 
medium/high spin i.e. the range most relevant for GW observations.


\section{Concluding remarks}

In this work we have produced general formulae for the frequency and damping rate 
of the early ringdown signal of general relativistic rotating UCOs. These expressions are based on the 
eikonal approximation and the presence of an unstable circular photon orbit in the UCO's exterior spacetime  
and parametrise the deviation from a Kerr black hole ringdown in terms of the body's multipolar structure. 
As an example, we have considered a thin-shell gravastar model and quantified 
its ringdown properties as a function of its compactness and spin. 

Our results clearly suggest that GW ringdown observations moderately more accurate than 
the ``champion" detection so far (GW150914) could begin putting to the test (at least) some of the 
proposed UCOs as alternatives to black holes. 

Our scheme should be viewed as complementary to ongoing efforts to 
model the late time $w$-mode/echoes portion of UCO ringdowns  \cite{Mark:2017zmc,Maselli_etal2017}.
This could be particularly important in the event that these echoes are too weak to be identified 
in the GW signal and instead one would have to rely on the much stronger early ringdown for making
a reliable identification of the post-merger remnant.

It is also conceivable that the ringdown signal could receive contributions from the UCO's fluid modes
(e.g. the $f$-mode) -- a possibility that neither our scheme nor any other in the topical literature has been designed to face.  
However, a signal of this kind is expected to look very different from a Kerr ringdown and therefore will be a smoking gun
proof of the presence of an UCO.

Armed with the appropriate ringdown templates, the ultimate goal would be to carry out  ``spacetime mapping", 
i.e. constrain the deviation of the multipolar structure of compact objects from the 
known moments of Kerr black holes (an idea first put forward in the context of extreme-mass-ratio 
inspirals as GW sources for LISA~\cite{Ryan:1995}).

Constructing rigorous match-filtering ringdown templates for an arbitrary UCO is not feasible yet 
but as a first stab at the problem one could combine the numerical fit formula for the exact fundamental 
Kerr QNM/ringdown (the ``offset function''  of Ref.~\cite{Glampedakis:2017}) with the 
$\mathcal{O} (\Omega^2)$ and $\mathcal{O} (\Omega^3)$ multipolar correction terms of our 
Eqs.\,(\ref{omR}), (\ref{omI}). A template of this type would be an ideal tool for 
observationally constraining the post-merger spacetime's deviation from Kerr.
The waveform's overall accuracy could be enhanced by repeating the 
present calculation at $\mathcal{O} (\Omega^4)$ accuracy using the formalism of Ref.~\cite{Yagi_etal:2014}. 
Furthermore, we could broaden our understanding of possible deviations by considering 
more examples of rotating UCOs, such as boson stars \cite{Ryan1997PhysRevD} or anisotropic 
stars \cite{BowersLiang1974ApJ}.

Of course, this ambitious programme would also have to deal with data analysis issues 
\cite{Thrane:2017ll,Baibhav_etal:2017} and/or with UCOs  that look very much like black holes 
(e.g. black holes endowed with a quantum ``firewall" structure at the horizon~\cite{Almheiri:2013}, although the 
multipoles of these systems with rotation remain to be determined and their ringdown is expected to contain late time echoes).
Some of these issues will be addressed in a future publication.  

\acknowledgements

We thank Emanuele Berti for valuable feedback on an early draft of this work
and acknowledge networking support by the COST Action GWverse CA16104.


\bibliography{biblio.bib}

\end{document}